# Texas A&M Penning Trap Facility – Design of the Measurement Trap

M. Mehlman[a], D. Melconian[a], P.D. Shidling[a]

[a]Cyclotron Institute, Texas A&M University, 3366 TAMU, College Station, TX, 77843-3366.

**Abstract.** A tandem Penning trap facility has been designed and is under construction at the Texas A&M University Cyclotron Institute (TAMU-TRAP). The initial experimental program will be the study of correlation parameters for T=2 superallowed beta-delayed proton emitters. The measurement trap is a unique large-bore optimized 5-electrode cylindrical Penning trap, which features a 90 mm free radius, larger than in any existing Penning trap. This novel geometry allows for full radial containment of decay products of interest. The trap has also been designed to exhibit a "tunable" and "orthogonalized" geometry, which is useful for alternate experiments.

## INTRODUCTION

The primary scientific goals of TAMU-TRAP are three fold: to improve the understanding of isospin mixing corrections, to provide new nuclei from which to extract measurements of $V_{ud}$, and to search for the existence of scalar currents in the weak interaction, all via precision beta-decay experiments. A Penning trap system, which provides a low-energy, point-like source of ions, is particularly well suited to such measurements since the magnetic field used for radial confinement of the ions can simultaneously serve to contain a wide array of charged decay products in an appropriately designed geometry. The first experiment will be the measurement of the beta-neutrino correlation parameter for T=2 nuclides, starting with $^{32}$Ar (as measured by Adelberger et. al [1]), and continuing with $^{20}$Mg, $^{24}$Si, $^{28}$S, $^{36}$Ca, $^{40}$Ti, and $^{48}$Fe. In addition to these initial experiments, the TAMU-TRAP facility is planned to enable future nuclear physics inquiry through a flexible, well-characterized geometry that will be suitable for a wide range of measurements, in particular those that require a good quadrupole field within the trapping volume.

## DESIGN OF TAMU-TRAP

In order to be suitable for precision beta-decay experiments, the trap geometry must: a) provide a cold, spatially localized source of ions with minimal spread in energy; b) allow for full containment of decay products of interest; c) provide for disk-shaped detectors of arbitrary voltage to be placed at either end of the trap in order to detect decay products. For additional nuclear physics experiments it is useful for the trap to display a good quadrupole electric field (i.e., be "tunable") and to be "orthogonalized" in order to allow for the trapping potential to be tuned mid-measurement, as discussed by Gabrielse et. al [2]. Real-world constraints are imposed by the length and bore size of the Agilent 7T solenoidal magnet that will contain the trap in addition to machining and assembly considerations.

No existing trap design could be scaled or modified to satisfy each of these stipulations, so a new large-bore, short-endcap Penning trap was characterized from first principles. Following in part the discussion in Gabrielse et. al [2], a full solution for the electric field of such a trap was derived and used to optimize the electrode configuration. The resulting design can be seen in Fig. 1. Key features are the 90 mm free radius, larger than in any existing Penning trap, and the flat endcap electrodes, which are capable of being placed at an arbitrary potential. Physical considerations, such as gaps between electrodes, have also been taken into account. The geometry as optimized is well suited to precision beta-decay experiments, and will also be useful for a wide range of future nuclear physics studies due to the well-understood nature of the resulting electric field.

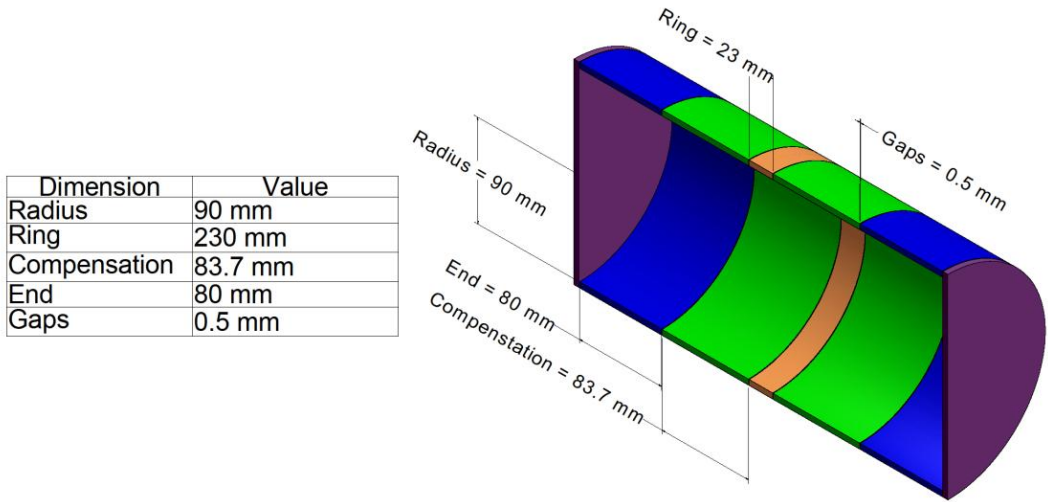

| Dimension | Value |
|---|---|
| Radius | 90 mm |
| Ring | 230 mm |
| Compensation | 83.7 mm |
| End | 80 mm |
| Gaps | 0.5 mm |

**FIGURE 1.** Optimized geometry for the TAMUTRAP measurement trap.

## SIMULATION

SIMION [3], an ion trajectory and field simulation program, was used as a check on the analytically derived electric field solution. The geometry was modeled within SIMION at an enlarged scale of 100 times the actual configuration in order to minimize simulation error due to finite grid size. Table 1 shows the expansion coefficients of the electric field when expanded in Legendre polynomials around the trap center according to both the analytic and simulated solutions for the geometry shown in Fig. 1. Apart from the constant offset term, $C_0$, the expansion is dominated by the quadrupole term, $C_2$, as desired. Odd coefficients (not shown), which should be 0 by symmetry across z=0, are smaller than $10^{-16}$ as calculated by both the analytic solution and SIMION simulation (non-zero odd analytic terms are the result of machine precision implementations of Legendre polynomials and Bessel functions and truncated sums).

**TABLE 1.** Expansion coefficients for the proposed geometry.

| Expansion Term | Analytic Solution | Simulation |
|---|---|---|
| $C_0$ | $-5.04 \times 10^{-1}$ | $-5.02 \times 10^{-1}$ |
| $C_2$ | $+5.12 \times 10^{-1}$ | $+5.13 \times 10^{-1}$ |
| $C_4$ | $+3.25 \times 10^{-6}$ | $+2.29 \times 10^{-3}$ |
| $C_6$ | $+4.24 \times 10^{-2}$ | $+3.73 \times 10^{-2}$ |
| $C_8$ | $-7.98 \times 10^{-2}$ | $-8.04 \times 10^{-2}$ |

## CONCLUSIONS AND OUTLOOK

Geometrical optimization of the TAMU-TRAP measurement Penning trap has been completed. An analytic solution for a new short-endcap Penning trap employing an open geometry suitable for both low-energy precision tests of the SM and additional nuclear physics measurements was derived from first principles. The electric field of the resulting structure has been described analytically and was designed to be both "tunable" and "orthogonalized", making the geometry attractive to a wide range of nuclear physics experiments. To confirm the validity of the characterization of the electric field, the geometry was modeled in SIMION and compared to the analytic solution. The calculated and simulated solutions agree to a degree acceptable considering the limitations imposed on SIMION by RAM and processing availability.

## ACKNOWLEDGMENTS

The authors would like to acknowledge the United States Department of Energy and the Texas A&M University Cyclotron Institute for funding this work. Additional thanks go to the CPT group at ANL for advice and guidance, as well as the faculty and staff of the Cyclotron Institute.